\documentclass[aps,prx,showpacs,superscriptaddress,twocolumn]{revtex4-2}

\usepackage{amsmath}
\usepackage{amssymb}
\usepackage{xcolor}
\usepackage{graphicx}
\usepackage{dcolumn}
\usepackage{bm}
\usepackage{natbib}
\usepackage{float}
\usepackage{ulem}
\begin{document}

	\title{Parallel assembly of arbitrary defect-free atom arrays with a multi-tweezer algorithm} 

	\author{Weikun Tian}
	\affiliation{Centre for Quantum Technologies, National University of Singapore, 117543 Singapore, Singapore}
	\author{Wen Jun Wee}
	\affiliation{Centre for Quantum Technologies, National University of Singapore, 117543 Singapore, Singapore}
	\author{An Qu}
	\affiliation{Centre for Quantum Technologies, National University of Singapore, 117543 Singapore, Singapore}
	\author{Billy Jun Ming Lim}
	\affiliation{Centre for Quantum Technologies, National University of Singapore, 117543 Singapore, Singapore}
	\author{Prithvi Raj Datla}
	\affiliation{Department of Physics, National University of Singapore, 117542 Singapore, Singapore}
	\author{Vanessa Pei Wen Koh}
	\affiliation{Centre for Quantum Technologies, National University of Singapore, 117543 Singapore, Singapore}
	\author{Huanqian Loh}
	\email[]{phylohh@nus.edu.sg}
	\affiliation{Centre for Quantum Technologies, National University of Singapore, 117543 Singapore, Singapore}
	\affiliation{Department of Physics, National University of Singapore, 117542 Singapore, Singapore}
	
    \date{\today}

	\begin{abstract}
	    Defect-free atom arrays are an important precursor for quantum information processing and quantum simulation. Yet, large-scale defect-free atom arrays can be challenging to realize, due to the losses encountered when rearranging stochastically loaded atoms to achieve a desired target array. Here, we demonstrate a novel parallel rearrangement algorithm that uses multiple mobile tweezers to independently sort and compress atom arrays in a way that naturally avoids atom collisions. With a high degree of parallelism, our algorithm offers a reduced move complexity compared to both single-tweezer algorithms and existing multi-tweezer algorithms. We further determine the optimal degree of parallelism to be a balance between an algorithmic speedup and multi-tweezer inhomogeneity effects. The defect-free probability for a 225-atom array is demonstrated to be as high as 33(1)\% in a room temperature setup after multiple cycles of rearrangement. The algorithm presented here can be implemented for any target array geometry with an underlying periodic structure.
    \end{abstract}
    
\maketitle

\section{Introduction}
Individual neutral atoms trapped in optical tweezer arrays with programmable geometries and interactions have become a powerful platform for quantum simulation \citep{Lukin_Ebadi_2021_256atomsimulator, Lahaye_Browaeys_2020_rydbergreview, 
Browaeys_Scholl_2021_triangle, Lukin_Semeghini_2021_spinliquid, 
Browaeys_Scholl_2022_XXZmodel, 
Bakr_Spar_2022_Li6Fermion, Gross_Lorenz_2021_ramansideband_rydbergdressing, Whitlock_Wang_2020_DMD}, quantum computation \citep{Whitlock_Morgado_2021_review_rydbergsimulation,  You_Wu_2021_RydbertatomReview, Ni_Kaufman_2021_reviewtweezers, Biedermann_Jau_2016_entanglingAtomicSpins, Pritchard_Picken_2018_entanglement, Ahn_Kim_2022_MIS, Lukin_Bluvstein_2022_CoherentTransport, Saffman_Graham_2022_multiqubitentangle}, quantum metrology \citep{Endres_Madjarov_2019_tweezerclock,
Kaufman_Young_2020_Srclock}, and foundational studies of quantum mechanics \citep{Endres_Choi_2021_emergent}. Recent work has extended this platform beyond alkali Rydberg atoms to encompass alkaline earths \citep{Endres_Madjarov_2019_tweezerclock, Kaufman_Young_2020_Srclock, Endres_Choi_2021_emergent, Kaufman_Jenkins_2022_Yb,
Takahashi_Okuno_2022_Yb,
Thompson_Ma_2022_Yb}, mixed-species arrays \citep{Cornish_Brooks_2021_CsRb, Bernien_Singh_2022_dualspecies,
Zhan_Sheng_2022_mixedisotope}, and molecules \citep{Ni_Liu_2018_NaCs, Doyle_Anderegg_2019_CaF, Zhan_He_2020_moleculeformation, Cheuk_Holland_2022_bichromatic}, elevating the versatility of the tweezer array platform. 

Many of these quantum science applications would benefit from the ability to generate large-scale defect-free atom arrays \citep{Whaley_Weiss_2004_ZeroEntropy}. In quantum simulation studies of quantum magnetism, for instance, defect-free atom arrays are crucial for obtaining clean measurements of order parameters \citep{Browaeys_Scholl_2021_triangle}. A pioneering method for assembling defect-free atom arrays uses a sequence of state-dependent translations on an optical lattice to sort atoms in parallel \citep{Weiss_Kumar_2018_OpticalLatticeRearrangement}. For optical tweezer arrays, one can use an auxiliary set of mobile tweezers \citep{Browaeys_Barredo_2016_defectfree2D, Lukin_Ebadi_2021_256atomsimulator,
Zhan_Sheng_2021_efficientpreparation} to rearrange single atoms that are stochastically loaded \citep{Andersen_Grunzweig_2010_graymolassesloading,
Regal_Brown_2019_graymolassesloading,
Loh_Aliyu_2021_D1magicalwavelength,
Gadway_Ang_2022_graymolasses,
Kaufman_Jenkins_2022_Yb} in a base set of static tweezers. In this case, the filling fraction is typically limited by rearrangement loss from a finite transit time and imperfect transfer of atoms between the mobile and static tweezer sets. 

Of the two limitations, the first limitation can be mitigated by embedding the tweezer array setup in a cryogenic vacuum system \citep{Browaeys_Schymik_2022_insitu}. Using this method, the atom lifetime can be extended to 6000 seconds \citep{Lahaye_Schymik_2021_6000slifetime}, which is about two orders of magnitude higher than that for room temperature setups. A complementary method is to minimize the rearrangement time by reducing the number of moves required, which can be achieved with efficient atom-sorting algorithms. 

So far, efforts to search for optimal rearrangement algorithms have focused on single-tweezer movements, where only one atom is sorted at a time. These algorithms include the Linear Sum Assignment Problem (LSAP) solver, the compression algorithm \citep{Lahaye_Schymik_2020_Enhancedassembly}, and the heuristic cluster algorithm (HCA) \citep{Zhan_Sheng_2021_efficientpreparation}. The move complexity of these single-tweezer algorithms, however optimized, scale at best linearly with the target array size. New algorithms that can outperform the linear scaling limit would thus be more efficient for assembling large atom arrays. 

Recently, the first multi-tweezer assembly of defect-free atom arrays is reported in Ref.\ \citep{Lukin_Ebadi_2021_256atomsimulator}. However, the algorithm described therein requires more effort to avoid collisions, which limits its ability to reach full parallelism.

Here, we report the demonstration of a new rearrangement protocol that uses multiple tweezers to independently sort and compress atom arrays in parallel. Our algorithm avoids coupling among rows and among columns and naturally ensures collision-free moves. We quantitatively investigate the effects of our parallel sort-and-compression algorithm (PSCA) on reducing the rearrangement complexity and compare it against existing algorithms. After addressing additional experimental issues that need to be mitigated, we use the PSCA to achieve a high success probability $\left[ 33(1)\% \right]$ of assembling 225-atom defect-free triangle arrays in a room temperature setup. Finally, we apply the PSCA to a range of target array geometries encompassing triangular variants of systems that are interesting for studies of spin frustration and fractal physics \citep{Lukin_Semeghini_2021_spinliquid, Sachdev_Samajdar_2021_KagomeLatticeRydberg, Vishwanath_Verresen_2021_toriccode, Xu_Li_2022_FrustratedTriangle, Ng_Zhou_2017_rmpspinliquid, Xu_MyersonJain_2022_Sierpinski}.

\begin{figure*}[tb]
    \centering
    \includegraphics[width=0.95\textwidth]{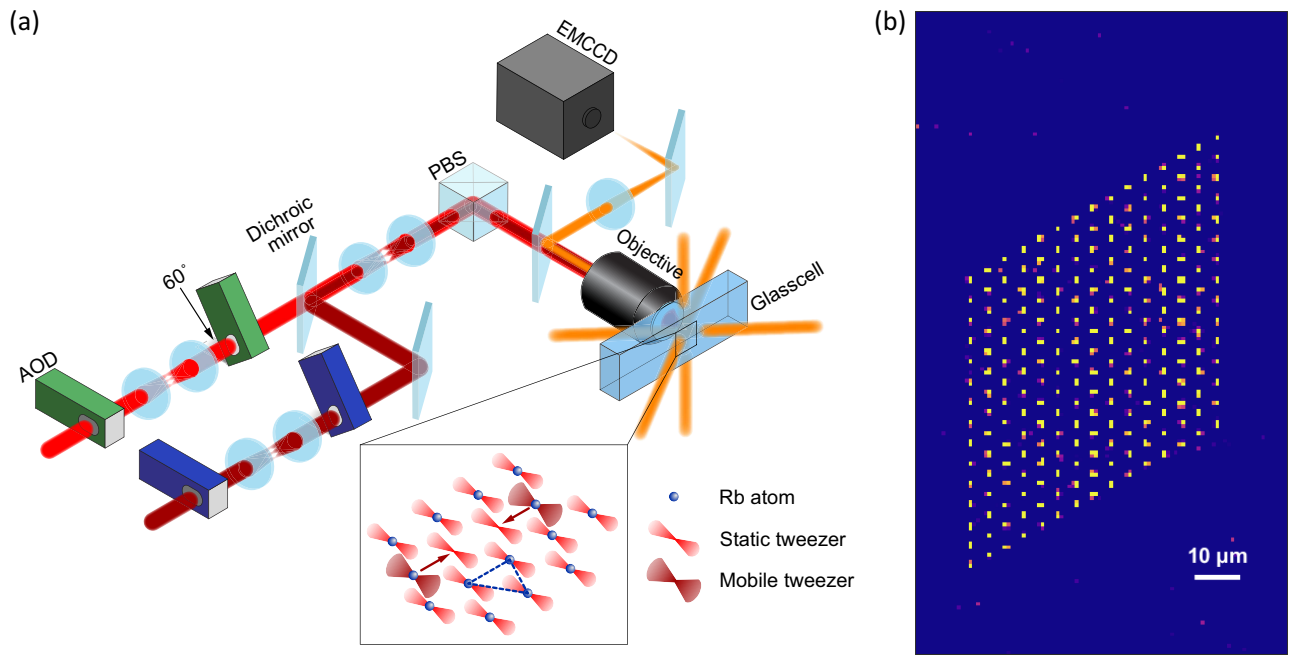}
    \caption{
    \textbf{Loading and rearrangement of Rb atoms.} 
    \textbf{(a)} The optical tweezer arrays for initial loading and rearrangement are generated by sending $808~\text{nm}$ (red) and $852~\text{nm}$ (dark red) laser beams through their respective pair of AODs. Each AOD pair has a $60^{\circ}$ relative orientation. (Inset) The pre-cooled cloud of Rb atoms are stochastically loaded into the static tweezers and rearranged with mobile tweezers into defect-free atom arrays. 
    \textbf{(b)} Single-shot fluorescence image of a defect-free triangular array with 225 atoms. 
}
    \label{fig1}
\end{figure*}

\section{Experimental Setup}
Our experiment sequence begins with loading laser-cooled $^{87}$Rb atoms into a base set of 20 $\times$ 20 static tweezers at 808~nm over a duration of 500~ms. The initial filling fraction typically ranges from 75\% to 78\% when $D_1$-enhanced gray molasses loading is used \citep{Andersen_Grunzweig_2010_graymolassesloading,
Regal_Brown_2019_graymolassesloading,
Loh_Aliyu_2021_D1magicalwavelength,
Gadway_Ang_2022_graymolasses,
Kaufman_Jenkins_2022_Yb}. To generate the static tweezers, an 808~nm beam is propagated through a pair of acousto-optic deflectors (AODs, AA Opto Electronic DTSXY-400-800), which are oriented at 60$^\circ$ relative to each other, before being focused down with a microscope objective (NA = 0.5) (Fig.~\ref{fig1}(a)). This setup allows us to generate a Talbot-free \citep{Saffman_Graham_2019_RydbergEntangleTalbot} equilateral triangular atom array. The static tweezer AODs are driven by software-defined radio to produce an array with an atomic spacing of 4.49(1)~$\mu$m. Using Stark shift measurements, we determine the average trap depth to be 300~$\mu$K with 4\% inhomogeneity (relative standard deviation) across the array. We characterize the 808~nm tweezer axial and radial trap frequencies to be 9.0(4)~kHz and 57(2)~kHz respectively, while the trap waist is 0.98(2)~$\mu$m.

To transform the randomly loaded atom array to a defect-free array, we first image the initial array by collecting fluorescence photons on an electron multiplying charge coupled device (EMCCD) camera in 45~ms. Each site of the tweezer array is pre-calibrated with a set of coordinates $\left\{ r \right\}_{i,j}$ with an associated region of interest \citep{Gerbier_Qu_2020_FluoresenceImaging} and a fluorescence threshold $T_{i,j}$ to determine if the site is loaded with an atom. This calibration converts the image of the array into a binary occupancy matrix that is then passed into our algorithm, which calculates a sequence of required moves for rearranging the atoms into a user-defined target pattern.

The physical moving of the atoms is performed by an auxiliary set of mobile tweezers that is spatially overlapped with the base tweezers. The mobile tweezers are formed by deflecting 852~nm light with another pair of AODs that are driven by a two-channel arbitrary waveform generator (AWG M4i.6631-x8, Spectrum Instrumentation). Individual waveforms for transporting an atom from one site to another are pre-calculated and stored in the random-access memory of a computer. The computer then picks, sums (if multiple atoms are involved in a single move), and concatenates the waveforms before loading them onto the AWG. Subsequently, the AODs deflect the mobile tweezer light accordingly. During this stage, $\Lambda$-enhanced gray molasses cooling is applied.

Typically, to transfer atoms from the static tweezers to the mobile tweezers, the latter is ramped up to a trap depth of 2.1~mK in 60~$\mu$s when up to 20 mobile tweezers are simultaneously turned on. At this trap depth, the axial and radial trap frequencies are 23.4(8)~kHz and 145(3)~kHz respectively, while the trap waist is 0.98(1)~$\mu$m. The atoms are then transported at a constant speed of 130~$\mu$m/ms, during which they are constrained to move along the rows and columns spanned by the static tweezers. The transport time is adiabatic compared to the static tweezer radial trap frequencies, thereby minimizing the periodic perturbations to the potential experienced by the transported atoms as they traverse the static tweezers. When the atoms have reached their target positions, the mobile tweezers are ramped down to zero in another 60~$\mu$s. At each joint of the ramp and transport, the phases of each waveform are matched, so that the atom in motion does not experience any sudden jumps in the mobile tweezer potential during rearrangement. 

Besides moving the atoms at a constant velocity, we attempted to move them with a constant jerk. However, compared to Ref.\ \citep{Lukin_Bluvstein_2022_CoherentTransport} where the constant jerk was implemented without the atoms in motion passing through other optical traps, we observe a lower rearrangement fidelity. We interpret the decrease in rearrangement fidelity to be a result of atoms passing through the static tweezers with a large velocity. This introduces a sudden change in potential experienced by the atoms, which can cause parametric heating. 

After the rearrangement, another fluorescence image is taken to verify the atom occupancy. Fig.~\ref{fig1}(b) shows a single-shot image of a defect-free triangular array with 225 atoms.

\section{Parallel sort-and-compression algorithm and results}
\subsection{Algorithm overview}
The rearrangement algorithm starts by identifying the center region of the base array as the target and its surrounding rows and columns as the reservoir, from which we withdraw atoms to fill the target array. 

\begin{figure}[tb]
    \centering
    \includegraphics[scale=1.07]{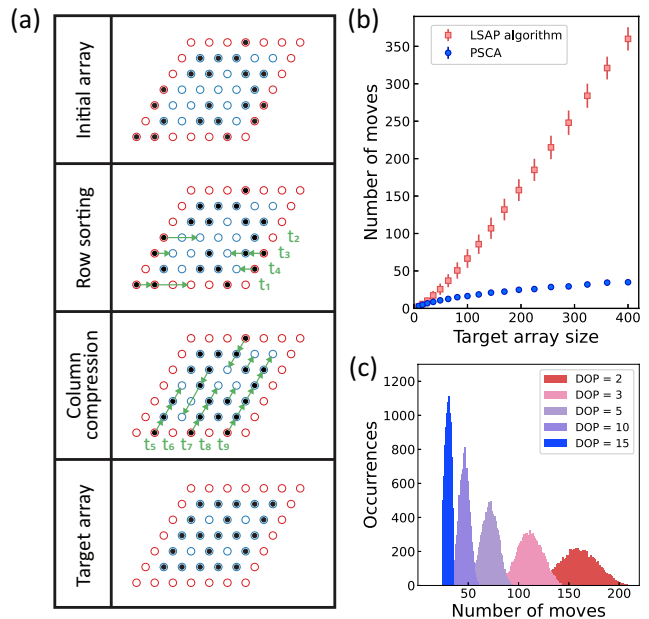}
    \caption{\textbf{Comparison of the multi-tweezer PSCA against the single-tweezer modified LSAP algorithm.}
    \textbf{(a)} Basic steps for the assembly of a defect-free kagome array using the PSCA. The red and blue open circles denote reservoir and target sites respectively. The black filled circles denote atoms, and the green arrows indicate the action of the mobile tweezers. The moves labeled by the same $t_{i}$ take place simultaneously.
    \textbf{(b)} Simulated scaling of the number of moves for different compact target array sizes using the PSCA (blue circles) and the modified LSAP algorithm (pink squares). The simulations are run with 10,000 iterations of different random initial loading configurations.
    \textbf{(c)} Simulated histogram of the number of moves to generate 15-by-15 compact target arrays with different DOPs.
}
    \label{fig2}
\end{figure}

The PSCA comprises two parts: row sorting and column compression. Due to the stochastic nature of loading single atoms into the base array, the initial number of atoms present in each column can vary. Further, for arbitrary geometries, the atoms needed in each column can be different, \textit{e.g.}\ for the kagome array, some columns are half-filled. The row sorting procedure redistributes the atoms among different columns such that each column has the same number of atoms as required in the target array. The row sorting is done row by row and each instance of row sorting aims to help the columns that are most in need of atoms. The choice of which rows to sort first can affect the efficiency of the row-sorting procedure. Here, we choose a heuristic that prioritizes rows that are closest to half-filled. We find that this heuristic performs better than prioritizing rows that are least or most filled, as half-filled rows have the best balance between the number of atoms that can support other columns and vacancies to redistribute to. The column compression then moves the atoms column-by-column into their target positions. Fig.~\ref{fig2}(a) shows an example of rearranging a stochastically loaded $7\times7$ array into a defect-free kagome array. 

For each case of row sorting (column compression), all the atoms move within the same row (column) independently of other rows (columns). Further, the order of the atoms is preserved after the rearrangement, \textit{i.e.}\ the atoms do not swap positions during the move. As a result, we avoid atom loss from collisions. 

\subsection{Reduced move complexity}

\subsubsection{Comparison with existing single-tweezer algorithms}
In real-time atom rearrangement, transporting the atoms typically takes an order of magnitude more time compared to calculating the moves \citep{Lahaye_Schymik_2020_Enhancedassembly}. Our algorithm requires up to 1.5~ms of calculation time for rearranging a 15 $\times$ 15 defect-free array from a 20 $\times$ 20 base array on a desktop computer with 32~GB of random-access memory. This is a small fraction compared to the total duration (60~ms) of our rearrangement stage. Therefore, minimizing the move complexity (\textit{i.e.}\ number of moves) is the primary goal of rearrangement algorithms.
 
Several single-tweezer rearrangement algorithms have already been developed. The search for the optimal atom-sorting protocol can be framed as solving the LSAP. LSAP algorithms use solvers (e.g.\ Jonker-Volganent algorithm) to minimize a cost function, which can be defined to be either the total travel distance $\sum_{\text{moves}\ i} \ell_i$ (conventional LSAP algorithm) or $\sum_{\text{moves}\ i} \ell_i^2$ (modified LSAP algorithm). The moves are then reordered to avoid collisions \citep{Lahaye_Schymik_2020_Enhancedassembly}. Alternatively, one can use the compression algorithm, which assembles the defect-free array layer by layer starting from the center region, or the HCA. The HCA first identifies the open regions (adjacent to reservoir sites) and the closed regions (isolated from reservoir sites) in the target array and removes the obstacles between them. The open regions then can be filled with reservoir atoms via a simple mapping \citep{Zhan_Sheng_2021_efficientpreparation}. For single-tweezer rearrangement algorithms, a key challenge comes from the fact that at most one vacancy in the target array can be filled with one move, thereby limiting the move complexity to scale as the target array size $N$ or larger.
 
The PSCA, in contrast, partitions the two-dimensional optimization problem into a number of one-dimensional optimization problems. The optimized one-dimensional moves can be carried out simultaneously, which takes much less time compared to single-tweezer algorithms where these moves need to be done sequentially.

To evaluate the efficiency of the PSCA, we compare the move complexity of our algorithm against the modified LSAP algorithm for a compact target array. The modified LSAP algorithm is chosen here for comparison as it has been previously demonstrated to be more efficient than both the conventional LSAP solver and greedy algorithms for generating compact arrays \citep{Lahaye_Schymik_2020_Enhancedassembly}. In our simulation, we solve for the rearrangement strategy using both algorithms applied to different random initial atom arrays with a loading probability per site of 75\%. Running the simulations multiple times for a given target array size $N$ spanned by $L$ rows and $L$ columns, where $L\times L = N$, yields a distribution of moves. We then vary $N$ to simulate the move complexity. Fig.~\ref{fig2}(b) shows the scaling of the number of moves for different target array sizes using the two algorithms, where the PSCA invokes up to 15 simultaneous mobile tweezers. With increasing target array size $N$, the average number of moves scales as $N^{1.16(1)}$ for the modified LSAP algorithm, whereas it only scales as $N^{0.48(2)}$ for the PSCA. 

\begin{figure}[tb]
    \centering
    \includegraphics[scale=1.07]{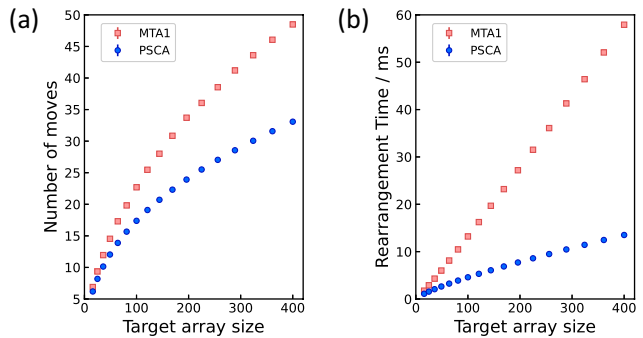}
    \caption{
    \textbf{Comparison of the PSCA against MTA1.}
    \textbf{(a)} Simulated scaling of the number of moves for different target array sizes using the PSCA (blue circles) and MTA1 (pink squares).
    \textbf{(b)} Simulated scaling of the atom rearrangement time for different target array sizes using the PSCA (blue circles) and MTA1 (pink squares). 10,000 iterations of the simulation are run with different random initial loading configurations. The error bars are smaller than the marker size on the plots.
    }
    \label{fig3}
\end{figure}

\subsubsection{Comparison with existing multi-tweezer algorithms}
To beat the linear scaling limit for move complexity, we turn to using multiple tweezers for atom rearrangement, of which the first protocol was developed and experimentally demonstrated in Ref.\ \citep{Lukin_Ebadi_2021_256atomsimulator} (herein referred to as MTA1). MTA1 begins with a pre-sorting stage to ensure that each column will have sufficient atoms to fill the target traps in that column. Subsequently, atoms are shuffled to their target traps via vertical scans. These stages appear to be similar to the row sorting and column compression steps of the PSCA, yet important differences exist between the two algorithms.

In the pre-sorting stage, to fill a deficient column, MTA1 decides which of the two sides of the column has surplus atoms. For each defect in that column, the algorithm then searches for the nearest atoms in the same row as the defect from the surplus side to fill the defect. However, there is no guarantee that these atoms come from columns that have an overall surplus. As a result, some moves are likely to be propagating defects through columns until the column with an atom surplus is reached. The PSCA avoids this redundancy in moves with its heuristic row-sorting strategy, which ensures the ability to shift the surplus atom from one column to any other column with one move.

We also note that MTA1 identifies the target and reservoir arrays such that the reservoir rows are interleaved as alternating rows between the target array rows. With such a geometry, up to two scans (one downward and one upward) may be required to complete rearranging one row. In contrast, the PSCA requires at most one compression operation to complete rearranging one column.

To evaluate the extent of parallelism, we compare the move complexity of the PSCA with that of MTA1. For both algorithms, the simulations are conducted with the geometry and initial loading probability (55\%) from Ref. \citep{Lukin_Ebadi_2021_256atomsimulator}. Fig.~\ref{fig3}(a) shows the scaling of the number of moves for different target array sizes using the two algorithms. We find that the average number of moves scales as $N^{0.47(1)}$ for MTA1 whereas it scales as $N^{0.388(2)}$ for PSCA.

The MTA1 also requires that scans are programmed to include a wait duration for atom capture/release at each row (column) by default, even when not required. These scans can also contain periods without any atom rearrangement. The PSCA avoids these additional wait times since at least one atom will be moving, being captured or released at any given time. We extend our simulation to compare the time taken by both algorithms to rearrange atom arrays with different sizes as shown in Fig. \ref{fig3}(b), where the moving speed and capture time from Ref.\ \citep{Lukin_Ebadi_2021_256atomsimulator} are used. For a target array size of 400 atoms, PSCA requires only a quarter of the atom-rearrangement time (excluding hardware delays) compared to that of MTA1.

\subsubsection{Tunable degree of parallelism}
As a multi-tweezer algorithm, the PSCA offers a tunable degree of parallelism (DOP), which we define to be a constraint on the maximum number of tweezers allowed to be turned on for each move. For example, if one parallel move includes moving seven atoms in a column, where $\text{DOP} = 4$, the rearrangement will be separated into two moves with four atoms and three atoms moved, respectively. Fig.~\ref{fig2}(c) shows a histogram of number of moves required for the rearrangement with different DOPs applied to obtain a 15 $\times$ 15 target array. The average number of moves required for rearrangement decreases significantly from $160.0(2)$ to $29.61(3)$ as the DOP increases from 2 to 15, indicating that a higher DOP is more efficient. 


\subsection{Mitigation of multi-tweezer inhomogeneities}

While multiple mobile tweezers can speed up the rearrangement process, they also require extra care to minimize the trap inhomogeneity during rearrangement, which can otherwise be detrimental to the atoms. We discuss below two sources of trap inhomogeneity: first, for a given number of tweezers, intermodulation among different radiofrequency (RF) tones sent to the AODs can generate an array with unequal trap depths \citep{Lukin_Endres_2016_1D_assembly}; second, the average trap depth depends on the number of mobile tweezers that are simultaneously turned on.

\textit{RF intermodulation:} The generation of multiple tweezers requires sending multiple RF tones to the AODs. Nonlinearities in the AODs and RF amplifiers cause intermodulation (frequency mixing) among the different tones, which becomes stronger when more tones are sent to the AODs and when their phases are strongly correlated \citep{Lukin_Endres_2016_1D_assembly}. To mitigate the intermodulation, we randomize the initial phases assigned to the rearrangement waveforms \citep{Ni_Zhang_2022_NaCsArray}. Using these initial phases, we have achieved an inhomogeneity (relative standard deviation) of 2\% within a row or column of 20 mobile tweezers at fixed array positions by adjusting the amplitudes of the different RF tones. These optimized amplitudes are henceforth considered `preassigned' during rearrangement.

In theory, the phases can be further optimized to suppress the intermodulation and achieve better homogeneity if the required individual waveforms that need to be summed are known and static. However, the optimization cannot be carried out in real time when the frequencies are swept. Thus, we randomly assign a list of initial phases to the individual waveforms in advance to enable real-time AOD control. 

\textit{Tweezer-number-dependent trap depth:} During rearrangement, the number of mobile tweezers turned on can vary depending on the number of vacancies present in a row or column. Since the initial array loading is stochastic, the number of mobile tweezers required for different moves is also probabilistic. 

To generate a given number of mobile tweezers, we turn on a subset of the RF tones with preassigned amplitudes. We observe that as we simultaneously turn on more tones, the average mobile trap depth decreases, leading to greater losses during atom transportation. In principle, one can increase the overall tweezer power to compensate for the drop in average trap depth. However, too high a trap depth can lead to losses arising from nonadiabatic transfers between the mobile and static tweezers \citep{Browaeys_Barredo_2016_defectfree2D}. In other words, increasing the average trap depth does not eliminate losses, because its improvement of the rearrangement fidelity for moves involving many mobile tweezers comes at a cost of lower fidelity for few tweezers. Compensating for the inhomogeneous trap depths among different moves, which is exacerbated with a higher DOP, is thus nontrivial.

\textit{Other potential issues:} We also investigate other possible sources of heating arising from the tweezer arrays, such as parametric heating from intensity noise fluctuations of the static and mobile tweezers, as well as the effect of mobile tweezers on atoms in neighboring static traps. These contributions are found to be negligible (Appendix A).

\subsection{Optimal degree of parallelism}

\begin{figure}[tb]
    \centering
    \includegraphics[scale=0.48]{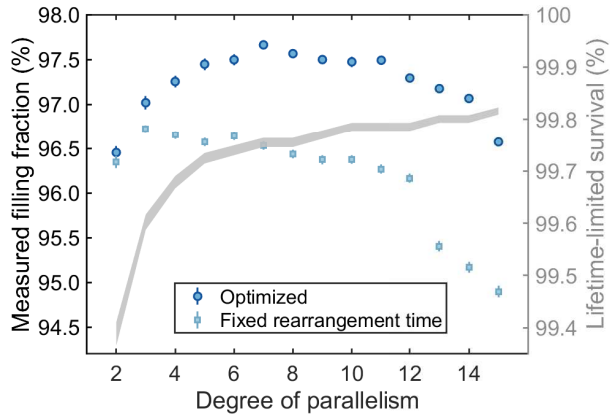}
    \caption{\textbf{Measured filling fractions achieved with different degrees of parallelism for a 15 $\times$ 15 compact array after one rearrangement cycle.} An algorithm with higher DOP performs the rearrangement more quickly, which increases the survival probability of the atoms given the trap lifetime decay (gray curve, $y$ axis on right, calculated with a pre-calibrated trap lifetime $\tau=33(1)$~s). However, the higher DOP also leads to more intermodulation, which causes intensity fluctuations that heat up the atoms (light blue squares, data taken with fixed rearrangement time of 200~ms). The overall measured filling fraction, taken with optimized rearrangement time windows, is highest when there is a trade-off between the two effects (dark blue circles).}
    \label{fig4}
\end{figure}

\begin{figure}[htbp]
    \centering
    \includegraphics[scale=0.68]{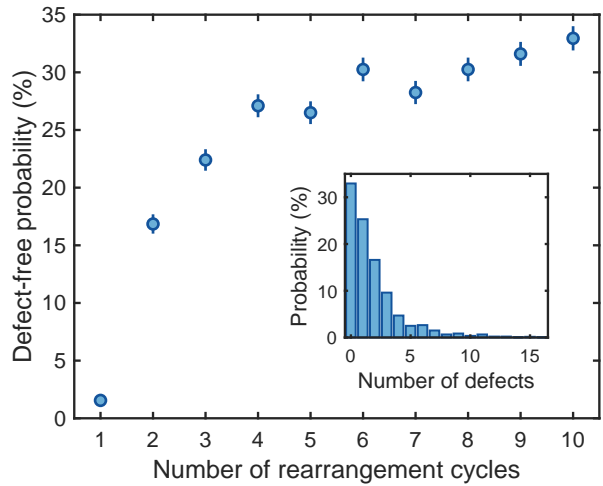}
    \caption{\textbf{Improvement of the defect-free probability with multiple cycles of rearrangement for a 225-atom compact target array (DOP = 7).}
    The defect-free probability increased to 33(1)\% after 10 cycles. The error bars correspond to $1\sigma$ equal-tailed Jeffrey’s prior confidence intervals. (Inset) Measured probability distribution of defects after 10 rearrangement cycles.}
    \label{fig5}
\end{figure}

A higher DOP yields a faster rearrangement, which improves the survival probability of an atom with a finite trap lifetime. On the other hand, a higher DOP can reduce the rearrangement fidelity by introducing more inhomogeneity in the mobile tweezer trap depths. Hence, we seek to find an optimal DOP by balancing the two effects against each other.

Of the two effects, we first consider the speedup in rearrangement time. The actual time taken for rearrangement varies from shot to shot and depends on how much the stochastically loaded initial array deviates from the target. In our experiment control sequence, we reserve a time window ranging from 60~ms to 200~ms, within which the rearrangement can be completed for most of the shots. Higher DOPs are more efficient and afford shorter time windows, which lead to higher atom survival probabilities in the presence of a finite trap lifetime (gray curve in Fig.~\ref{fig4}). 

On the other hand, the reduced rearrangement fidelity from a higher DOP can overwhelm the benefit of a speedup. To isolate the effect of multi-tweezer inhomogeneity, we set the rearrangement time window for all DOPs to be 200~ms, which gives a constant lifetime-related loss that can be factored out. Fig.~\ref{fig4} shows that the filling fraction in a 15 $\times$ 15 compact target array decreases as the DOP increases from 3 to 15 (light blue squares), which can be attributed to the increased inhomogeneity in mobile tweezer trap depths. The low filling fraction with $\text{DOP}=2$ is an exception and comes from the fact that the variance of the rearrangement time is so large that 0.8\% of the shots cannot be rearranged within the 200~ms time window. We further postulate that from DOP = 12 to DOP = 13, the mobile tweezer trap depth could have decreased below a threshold for adiabatic atom transport, such that the rearrangement fidelity is more significantly affected for DOP~$\geq$~13.

Combining these two effects, we optimized the rearrangement time window for different DOPs used to form the same 15 $\times$ 15 target array. The measured filling fractions with optimized time windows are shown as dark blue circles in Fig.~\ref{fig4}. The filling fraction increases initially, boosted by the reduced complexity of the PSCA, before declining at high DOP where the intermodulation effects dominate. At the optimal point (DOP = 7), we achieve a measured filling fraction of $97.66(4)\%$.

\subsection{Multi-cycle rearrangement}

Rearrangement losses from finite transit duration, tweezer intensity fluctuations, and imperfect capture and release of atoms from the mobile tweezers can introduce defects in the rearranged array. An effective solution is to use multiple rearrangement cycles to correct for defects arising from the previous cycles \citep{Birkl_Ohl_de_Mello_2019_multicycle}. These subsequent cycles require fewer moves and shorter rearrangement time, thereby leading to lower loss and higher filling fractions.

\begin{figure*}[tb]
    \centering
    \includegraphics[scale=0.52,trim={6cm 2cm 4cm 0}]{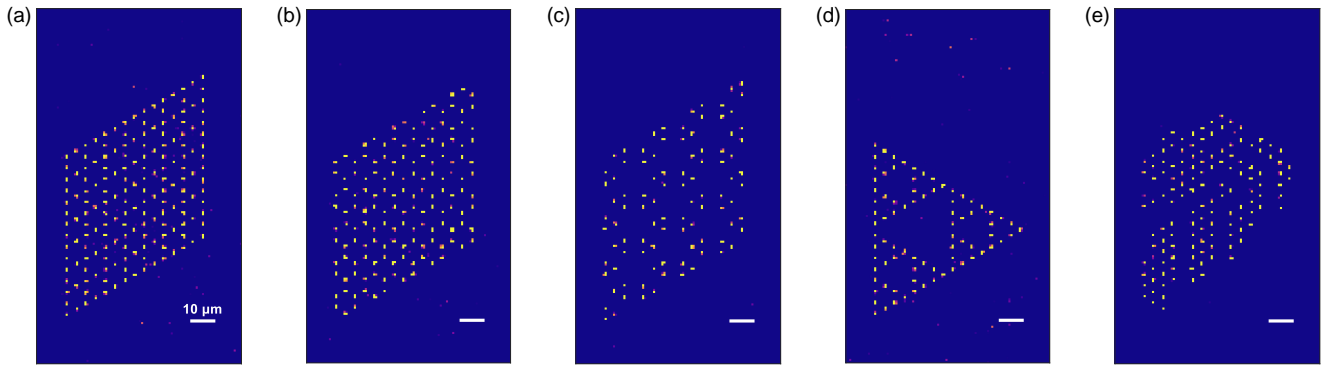}
    \caption{
        \textbf{Gallery of single-shot images of defect-free atom arrays with arbitrary geometries:}
        \textbf{(a)} kagome,
        \textbf{(b)} honeycomb,
        \textbf{(c)} link-kagome,
        \textbf{(d)} Sierpinski triangle, and
        \textbf{(e)} lion head symbol, an icon of Singapore.
        All scale bars indicate a distance of $10~\mu$m in the atom plane.
}
    \label{fig6}
\end{figure*}

Fig.~\ref{fig5} shows the 225-atom array defect-free probability increasing as more rearrangement cycles are used, which corresponds to fewer defects that remain to be corrected over time. Our target geometry is a compact triangular array and we set $\text{DOP}=7$ as optimized in the previous section. We find that the defect-free array probability improves from $1.5(3)\%$ to $16.9(8)\%$ after a second rearrangement cycle. After 10 rearrangement cycles, the defect-free probability stabilizes at 33(1)\%, which is an order-of-magnitude increase compared to one round of rearrangement while the experiment cycle time only increases by a factor of 1.8. The measured probability distribution of defects after 10 rearrangement cycles is shown as an inset in Fig.~\ref{fig5}. 

The defect-free array probability increases the most significantly at the second rearrangement cycle. This comes from the fact that the first rearrangement involves moving almost all the atoms as the initially loaded array is sparse. Given that the rearrangement losses scale as the number of atoms being moved, the first cycle results in a relatively low defect-free probability. For subsequent cycles, the arrays that need to be rearranged typically have a filling fraction over 97\% and only a few atoms (2.7 atoms on average) need to be moved. As a result, the rearrangement loss is exponentially lower, which leads to a significant increase in the defect-free probability.

Besides rearrangement losses, the other factors that limit the measured defect-free array probability include false detection of the atom occupation (mean detection fidelity 99.946(7)\%) and heating from atom fluorescence imaging (survival probability of $99.55(1)\%$ after one round of imaging). These are the same factors that set the upper bound on the defect-free array probability as the number of rearrangement cycles increases.

\subsection{Arbitrary geometries}

Besides compact triangle arrays, the PSCA can also generate any user-defined arbitrary geometry with an underlying periodic structure. Fig.~\ref{fig6} shows single-shot example images of defect-free atom arrays for the kagome, honeycomb, link-kagome, and Sierpinski triangle geometries, as well as the lion head symbol, which is an icon of Singapore.

\section{Conclusion}
In this paper, we have proposed and demonstrated a new parallel rearrangement algorithm for defect-free atom array assembly with an adjustable degree of parallelism corresponding to a constraint on the maximum number of mobile tweezers. For a 225-atom target array and DOP = 15, our parallel rearrangement algorithm significantly reduces the average number of moves to scale as the square root of the target array size. We have used our algorithm to experimentally realize large-scale defect-free arrays with hundreds of atoms with a high success probability up to $33(1)\%$ in a room temperature environment. Future work involving a detailed fine-tuning of the mobile tweezer trap depths and ramp speeds, bearing in mind their tradeoffs against the total rearrangement time, can potentially yield a higher optimal DOP, thereby further increasing the rearrangement parallelism.

Our rearrangement algorithm can be applied or adapted to most atom-array setups including mixed-species and molecule systems. Moreover, the results reported here are obtained with tapered amplifiers as the laser sources for both the static and mobile tweezers. We expect the defect-free success probability to be further improved with quieter and more powerful laser sources like titanium sapphire ring lasers. Such scaling up of atom arrays holds exciting possibilities for explorations of exotic quantum phenomena and for achieving higher quantum computation power.

\section*{Acknowledgments}
This research is supported by the National Research Foundation, Singapore and A*STAR under its Quantum Engineering Program (NRF2021-QEP2-02-P09); National Research Foundation (NRF) Singapore, under its NRF fellowship scheme (NRF‐NRFF2018-02); Singapore Ministry of Education Academic Research Fund Tier 1 (A-0004210-00-00), Singapore. We also acknowledge support from the Research Centres of Excellence program (NRF Singapore and the Ministry of Education, Singapore).

\section*{Appendix}

\subsection{Tweezer intensity noise}
Tweezer intensity noise, particularly that at twice the trap frequency, can be detrimental to the atoms by parametrically heating them out of the trap. We investigate the effects of intensity noise from both the static and mobile tweezer arrays by characterizing their respective intensity noise spectra.

To measure the noise experienced by atoms trapped in the static tweezer array, we generate a horizontal row of 20 tweezers at 300~$\mu$K trap depth and clip the array at the focal plane of a lens, such that only a single tweezer beam is sent onto a photodiode. The intensity noise at twice the radial trap frequency is measured to be -107.5(2)~dBc/Hz, which corresponds to a heating rate of $\Gamma = 0.57(7)~$Hz \citep{Thomas_Gehm_1998_ParametricHeating}. 

Similarly, we monitor the intensity noise of a mobile tweezer as it moves repeatedly across a distance of one site at the typical atom-transport speed (130~$\mu$m/ms). Here, the intensity noise at twice the radial trap frequency is -113.7(3) dBc/Hz and the corresponding heating rate for the 2.1~mK trap depth is $0.9(1)$~Hz.

For the above heating rates, we estimate the trap lifetime to be 3--5~s without any cooling of atoms. However, in the presence of $D_1$ cooling lasers, the atoms in the static tweezer array are measured to have a much longer trap lifetime (33(1)~s), indicating that parametric heating from tweezer intensity noise is well suppressed during rearrangement.

In addition, we monitor the heating effects on an atom trapped in a static tweezer while a nearby mobile tweezer is activated to move along a column adjacent to the static tweezer of interest. Compared to the case without any mobile tweezers turned on, the intensity noise at twice the static tweezer radial trap frequency is slightly higher at -103.8(3)~dBc/Hz, giving a heating rate of 1.4(2)~Hz. Nonetheless, we do not observe any statistical increase in the atom temperature before and after the mobile tweezer is activated, indicating that the heating rate from the neighboring mobile tweezer is negligible. 

%

\end{document}